\documentclass[aps,preprint]{revtex4}
\usepackage{amsmath}
\usepackage{amssymb}
\usepackage{tikz}
\usepackage{graphicx}
\usepackage{dcolumn}
\usepackage{bm}
\usepackage{epsfig}
\usepackage{psfrag}
\usepackage{amsfonts}
\usepackage{color}
\usepackage{gensymb}
\DeclareMathOperator\erf{erf}
\begin{document}
\draft

\title{\bf The edge switch transformation in microwave networks}

\author{ Vitalii Yunko, Ma{\l}gorzata Bia{\l}ous, and Leszek Sirko}

\address{Institute of Physics, Polish Academy of Sciences, Aleja Lotnik\'{o}w 32/46, 02-668 Warsaw, Poland\\
}

\date{\today}

\bigskip

\begin{abstract}

We investigated  the spectra  of resonances of  four-vertex microwave networks simulating both quantum graphs with preserved and with partially violated time-reversal invariance before and after an edge switch operation.
We show experimentally that under the edge switch operation the spectra of the microwave networks with preserved time reversal symmetry are level-1 interlaced, i.e., $\nu_{n-r}\leq \tilde \nu_{n}\leq \nu_{n+r}$, where $r=1$, in agreement with the recent theoretical predictions of [M. Aizenman, H. Schanz, U. Smilansky, and S. Warzel, Acta Phys. Pol. A {\bf 132}, 1699 (2017)]. Here, we denote by $\{\nu_{n}\}_{n=1}^{\infty}$ and $\{\tilde \nu_{n}\}_{n=1}^{\infty}$ the spectra of microwave networks before and after the edge switch transformation.  We demonstrate that the experimental distribution $P(\Delta N)$  of the spectral shift $\Delta N$ is close to the theoretical one.
Furthermore, we show experimentally that in the case of the four-vertex networks with partially violated time reversal symmetry the spectra are level-1 interlaced.  Our experimental results are supplemented by the numerical calculations performed for quantum graphs with violated time-reversal symmetry. In this case the edge switch transformation also leads to the spectra which are level-1 interlaced. Moreover, we demonstrate that for microwave networks simulating graphs with violated time-reversal symmetry the experimental distribution $P(\Delta N)$  of the spectral shift $\Delta N$ agrees within the experimental uncertainly with the numerical one.

\end{abstract}

\bigskip
\maketitle
\section{Introduction}

Quantum graphs introduced by Linus Pauling \cite{Pauling1936}  are widely used in the studies of  quantum systems whose classical counterparts display chaotic behavior \cite{Kottos1997,Pluhar2014}. Spectral fluctuations in quantum graphs were extensively studied in Refs.~\cite{Schanz2001,Berkolaiko2003,Bolte2003} while the localization phenomena and wavefunction statistics were investigated in \cite{Keating2003,Gnutzmann2004}. Bohigas, Giannoni and Schmit conjecture \cite{Bohigas1984}  predicts that such chaotic systems can be modelled by appropriate Gaussian ensembles of the random matrix theory (RMT). In this approach the three symmetry classes are distinguished: the Gaussian orthogonal ensemble (GOE) and the Gaussian symplectic ensemble (GSE)  with time-reversal  invariance ($T$ invariance), characterized respectively by the symmetry indices $(\beta=1)$ and  $(\beta=4)$, and the Gaussian unitary ensemble (GUE) with broken time-reversal invariance $(\beta=2)$.

 An important role in the experimental investigations of chaotic systems plays wave simulators: microwave networks \cite{Hul2004} and billiards \cite{Stoeckmann90} simulating, respectively, quantum graphs and billiards.
 Microwave networks are the only systems which allow for the experimental simulation of quantum systems modeled by all three classical ensembles in RMT: with $T$ invariance belonging to the Gaussian orthogonal ensemble  \cite{Hul2004,Lawniczak2008,Hul2012,Dietz2017,Lawniczak2019} and the Gaussian symplectic ensemble \cite{Stockmann2016} as well as systems without  $T$  invariance belonging to the Gaussian unitary ensemble  \cite{Hul2004,Lawniczak2010,Allgaier2014,Bialous2016,Lawniczak2017,Lawniczak2019b}.
 Microwave billiards have been  successfully used for simulations of systems with  $T$ invariance \cite{Stoeckmann90,Sridhar,Richter,Sirko1997,Hlushchuk2000,Hemmady2005} along with the ones with violated \cite{So1995,Stoffregen1995,Wu1998} and partially violated \cite{Dietz2010,Dietz2019} $T$ invariance.

Microwave networks \cite{Hul2004,Lawniczak2008,Lawniczak2010,Hul2012,Bialous2016,Lawniczak2019,Lawniczak2019b} can simulate quantum (metric) graphs with chaotic dynamics \cite{Kottos1997} because both systems are characterized by the equivalent equations: the telegrapher's equation describing a microwave circuit and the one-dimensional Schr{\"o}dinger equation describing a quantum system.

A metric graph $\mathbb{G}=\mathcal{(V,E,L)}$ is a structure composed with $\mathcal{V}$=$\{v_{i}\}$ vertices, $\mathcal{E}$=$\{e_{i}\}$ edges and the finite total length $\mathcal{L}$. Quantum graphs are commonly used for studying quantum chaos \cite{Kottos1997,Kottos2003,Pakonski2003,Hul2004,Gnutzmann2006} and localization phenomena \cite{Kaplan2001,Schanz2003}. It is important to point out that quantum graphs which can be considered as idealization of physical networks have been for many years the subject of fundamental research in mathematical physics \cite{Exner1996,Kostrykin2004,Kuchment2005,Fabiano2016}. Similarly to quantum graphs, microwave networks are the unions of vertices connected by edges. They are constructed of $\{v_{i}\}$ microwave joints (vertices) and $\{e_{i}\}$ microwave coaxial cables (edges).
By attaching the leads to the vertices of the graphs, the originally closed compact systems are converted into the scattering open ones.

The average counting function of the number of eigenvalues with the absolute value of  the wave number $k=\sqrt{E}$ smaller than $R$, where $E$ denotes energy, satisfies the Weyl's law
\begin{equation}
  N_{av}(R) = \frac{\mathcal{L}}{\pi}R + \mathcal{O}(1)\,.\label{eq:asym}
\end{equation}
The function $\mathcal{O}(1)$ in the limit $R\rightarrow +\infty $ is bounded by a constant of order of 1.

Different transformations of a quantum graph, e.g., an edge switch and an edge swap have an impact on the spectrum \cite{Aizenman2017}. The first of these transformations does not necessary preserve a graph topology while in the second modification the topology of a graph is preserved.

If we denote by $\{E_{n}\}_{n=1}^{\infty}$ and $\{\tilde E_{n}\}_{n=1}^{\infty}$ the spectra of the graphs before and after the transformation we will call the spectrum $\{\tilde E_{n}\}_{n=1}^{\infty}$ level-r (degree-r) interlaced with that of the original one if $E_{n-r}\leq \tilde E_{n}\leq E_{n+r}$, where  $n>r$.
The aim of this experimental investigation is a comparison of the edge switch transformation for the networks with preserved and with violated time-reversal invariance. The edge switch transformation is realized by exchanging one pair of edges $\{e_{i}\}$ both adjacent  to a common vertex $\{v_{i}\}$.  Although the edge lengths and the Neumann vertex boundary conditions, often called standard boundary conditions, are unaltered, the impact of the edge switch modification of a microwave network on its spectra is observed. We  confirmed experimentally  that the edge switch transformation leads to the spectra level-1 interlaced.

The edge switch may lead to the change of the graph topology, but it does not have to. In our case the topology, the edge lengths and the boundary conditions are preserved.  The original spectrum of the network becomes affected by these operations leading to a new spectrum, hence respectively, the sequences $\nu_{n}$ and $\tilde \nu_{n}$ are expressed by relation

\begin{equation}
\label{Eq.1}
\nu_{n-r}\leq \tilde \nu_{n}\leq \nu_{n+r}
\end{equation}\

for any $n>r$. In such a case the spectra before and after the transformation are degree-r interlaced. A common feature of the transformations discussed in this paper is that they do not affect the average spectral density of graphs and networks $\rho_{av}=\frac{\mathcal{L}}{\pi}$. This follows from the Weyl's
asymptotic formula (\ref{eq:asym}) and the fact that the sum of edge lengths is preserved
under the transformations considered here. In the Ref. \cite{Aizenman2017} it was shown  that an edge switch transformation in the graphs with preserved time-reversal invariance is degree-1 interlaced $(r=1)$.

\section{Experiment}
\subsection{Networks simulating quantum graphs with preserved time-reversal invariance}

The edge transformations were experimentally investigated using four-vertex (tetrahedral) microwave networks with preserved time-reversal invariance. They were connected to the Agilent E8364B vector network analyzer (VNA) via the HP 85133-616 microwave flexible cable, see Fig.~1(a), forming  open systems. The resonances $\nu_n$  of open microwave networks were evaluated by measuring a single-port  scattering matrix $S(\nu)$  as  a function of microwave frequency $\nu$. Then  the real parts of the wave numbers  $k_n$ are directly related to the positions $\nu_n$ of the resonances  $\mathrm{Re\,}k_n=\frac{2\pi }{c}\nu_n$, where $c$ is the speed of light in the vacuum.

The effect of opening of microwave networks and graphs was carefully studied in the series of papers \cite{Bialous2016,Dietz2017,Lawniczak2019}. It was proved that in the case of an unbalanced vertex, when the number of internal connections to a vertex is bigger than the number of external leads connected to a vertex, one deals with the Weyl networks and graphs for which the number of resonances are predicted by the Weyl's law \cite{Lawniczak2019}. Therefore, one external lead connected to a vertex or a single connection to the microwave vector network analyzer does not change the number of resonances in the studied system. The Weyl networks and graphs with time reversal symmetry and with broken time reversal symmetry were comprehensively studied in the papers \cite{Dietz2017} and \cite{Bialous2016}. It was demonstrated experimentally that the short-range and long-range fluctuation properties in the spectra of the networks studied in terms of the nearest neighbor level spacing distribution, the levels variance, the spectral rigidity and the power spectrum even in the case of missing levels are very well described by the corresponding numerical results and the random matrix theory.
This confirms that the microwave networks are very well suited for the experimental investigations of the distributions of the spectral shift $P(\Delta N)$. Furthermore, in our experiment, similarly to the closed systems, the edge switch transformation does not change the openness of the networks because only internal edges are subjected to switching.

\begin{figure}
\includegraphics[width=18.cm, angle=0]{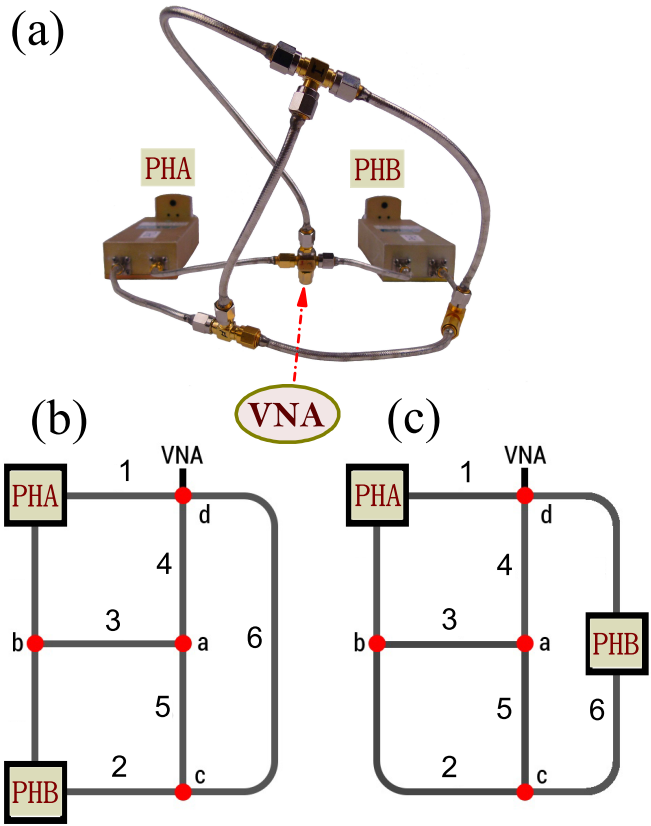}
\caption{ (a) Photo of a four-vertex (tetrahedral) microwave network containing two phase shifters PHA and PHB. The vertex connected to the vector network analyzer is marked with VNA. (b) The scheme of the four-vertex microwave network used for the creation of the first half of the configurations. (c) The  scheme of the four-vertex microwave network used in the creation of the second half of the configurations. } \label{Fig1}
\end{figure}

\begin{figure}
\includegraphics[width=12.cm, angle=0]{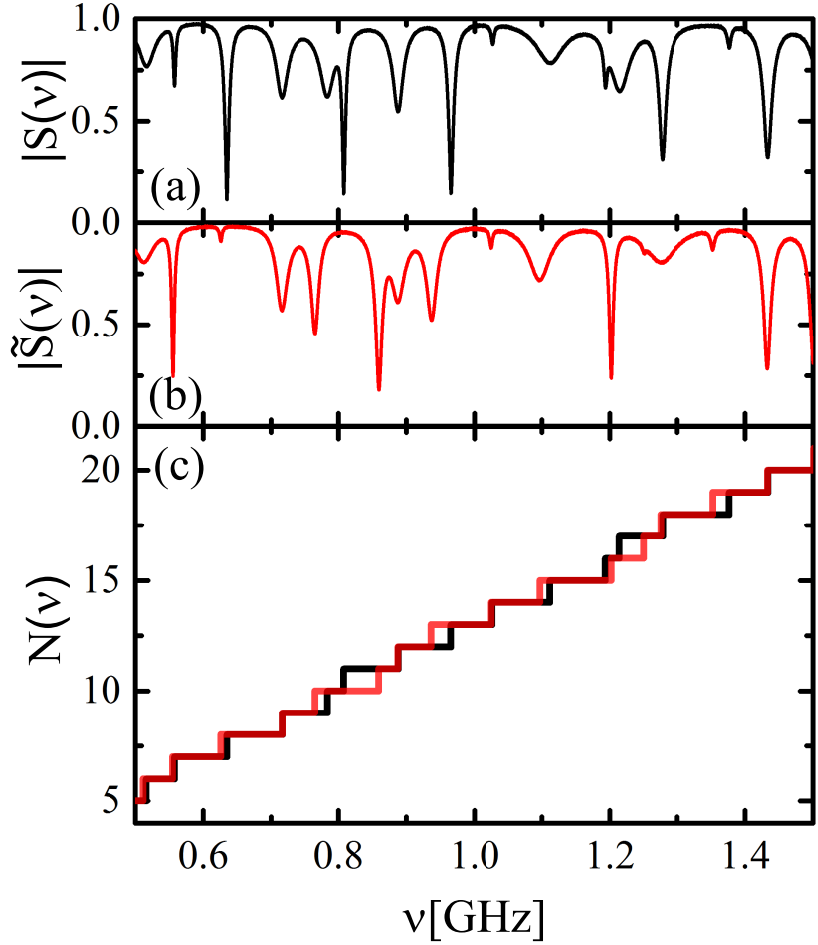}
\caption{(a) An example of the modulus of the scattering matrix $|S(\nu)|$ measured for the network with preserved time reversal symmetry before the edge switch transformation in the frequency range 0.5-1.5 GHz. (b) The modulus of the scattering matrix  $|\tilde S(\nu)|$  measured after the edge switch transformation. (c)  Comparison of the counting functions $N(\nu)$ (black lines) and $\tilde N(\nu)$  (red lines) of the ``original'' and ``switched'' configurations for the microwave network with preserved time reversal symmetry.} \label{Fig2}
\end{figure}

The fully connected tetrahedral networks used in the measurements were composed of bonds (coaxial cables), two phase shifters, three three-arm joints, and one four-arm joint connected to the VNA via a flexible microwave cable. The  coaxial cables (SMA-RG402) consist of an inner conductor of radius  $r_1=0.05$~cm and an outer concentric conductor of inner radius $r_2=0.15$~cm.  Both conductors are separated by dielectric material (Teflon) with the dielectric constant  $\varepsilon\simeq 2.06$.
The lengths of the two bonds were varied with the phase shifters (Advanced Technical Materials PNR P1507D) to obtain different realizations of the networks. The total optical length of the networks $L_{opt}=2.248\pm 0.008$~m was kept constant by increasing the length of one bond and decreasing the length of another one by the same amount.
The experiment was conducted in the following way. In the constructed network (Fig.~1(b) shows the scheme of the network) the length of the edge $1$ was increased from $0.697 \pm 0.001$ m to $0.747 \pm 0.001$ m by increasing the phase of the phase shifter $PHA$  from $0\degree$ to $60\degree$ in ten $6\degree$ steps, while the second phase shifter's $PHB$ phase in the edge $2$ was simultaneously decreased from $60\degree$ to $0\degree$ in the same number of steps. In this way the length of the edge $2$ was decreased from $0.612 \pm 0.001$ m to $0.562 \pm 0.001$ m. The scattering matrix $S(\nu)$ of the microwave networks was measured for all eleven phase shifters positions in the frequency window $\nu= 0.01-2.5$ GHz. For 2.5 GHz the maximum increase or decrease of the lengths of the edges $1$ and $2$ was equivalent to $0.6\lambda $, where $\lambda $ is the microwave wavelength in the coaxial cables.
In the frequency range $\nu= 0.01-2.5$ GHz we were able to identify first $36-37$ resonances. The number of the measured resonances  was in agreement with the one predicted by the Weyl's formula $N_{av}(\nu)=2L_{opt}\nu/c$, which in the frequency range 0.01-2.5 GHz yields $\simeq 37$ resonances. In order to avoid the missing resonances we analyzed the fluctuating part of the integrated spectral counting function $N_{fl}(\nu_i)=N(\nu_i)-N_{av}(\nu_i)$ \cite{Dietz2017}, that
is the difference of the number of identified eigenfrequencies
$N(\nu_i) = i$  for ordered frequencies $\nu_1 \leq \nu_2 \leq \ldots$ and
the number $N_{av}(\nu_i)$ predicted by the Weyl's formula  Eq.~(\ref{eq:asym}).
  In Fig.~2(a) we present an example of the modulus of the scattering matrix $|S(\nu)|$ (black full line) measured in the frequency range 0.5-1.5 GHz.
 By interchanging the coaxial cables $3$, length $0.170 \pm 0.001$ m, and $5$, length $0.243 \pm 0.001$ m, we modified the connectivity of these two edges which both terminate at a common vertex $a$. Namely, the cable $5$ originally connected to the vertices $a$ and $c$ is now connected to the vertices $a$ and $b$, while the cable $3$ which was originally connected to the vertices $a$ and $b$ is now connected to the vertices $a$ and $c$. Then, for the  ``switched'' networks we measured the scattering matrix $\tilde S(\nu)$  in the same frequency window and for the same positions of the phase shifters. In Fig.~2(b) we show the modulus of the scattering matrix $|\tilde S(\nu)|$ (red full line) measured in the frequency range 0.5-1.5 GHz. In order to increase the number of the network realizations, we placed the phase shifter PHB in the another position of the network (see Fig.~1(c)) and repeated the measurements. In this case the switch operation was realized by interchanging the cables $3$ and $2$ but keeping their connectivity in the vertex $b$. Here, the cables $2$ and $3$ are $0.327 \pm 0.001$ m and $ 0.170 \pm 0.001$ m long.
The pairs of ``original'' and ``switched'' configurations  (same phase shifter positions, both physically and optically) were analyzed in the following way. We compared the counting functions $N(\nu)$ and $\tilde N(\nu)$  of the ``original'' and ``switched'' configurations (see Fig.~2(c)) and calculated the distribution $P(\Delta N)$ of the spectral shift $\Delta N =N(\nu) - \tilde N(\nu)$ to be $\pm 1$ or $0$. In the case of the switch transformations we did not observe the instances of changing $\Delta N \ge 2$. It means that under the edge switch operation the spectra are level-1 interlaced with that of the original one.
In Fig.~3 we show the experimental distribution of the spectral shift $P(\Delta N)$  compared to the numerical results obtained for a tetrahedral graph analyzed in Ref. \cite{Aizenman2017}.  The experimental results in Fig.~3 are marked by the black bars while the numerical ones by the red ones.  The uncertainty of the spectral shift $P(\Delta N)$ was calculated as a standard deviation error obtained from 22 pairs of the measured spectra. The agreement between the experimental results and the numerical ones is very good. This can be explained by the fact that being tetrahedral graphs and networks both systems possess similar short- and long-range fluctuation properties in the spectra caused, i.a., by similar structures of periodic orbits \cite{Dietz2017}.

\begin{figure}
\includegraphics[width=15.cm, angle=0]{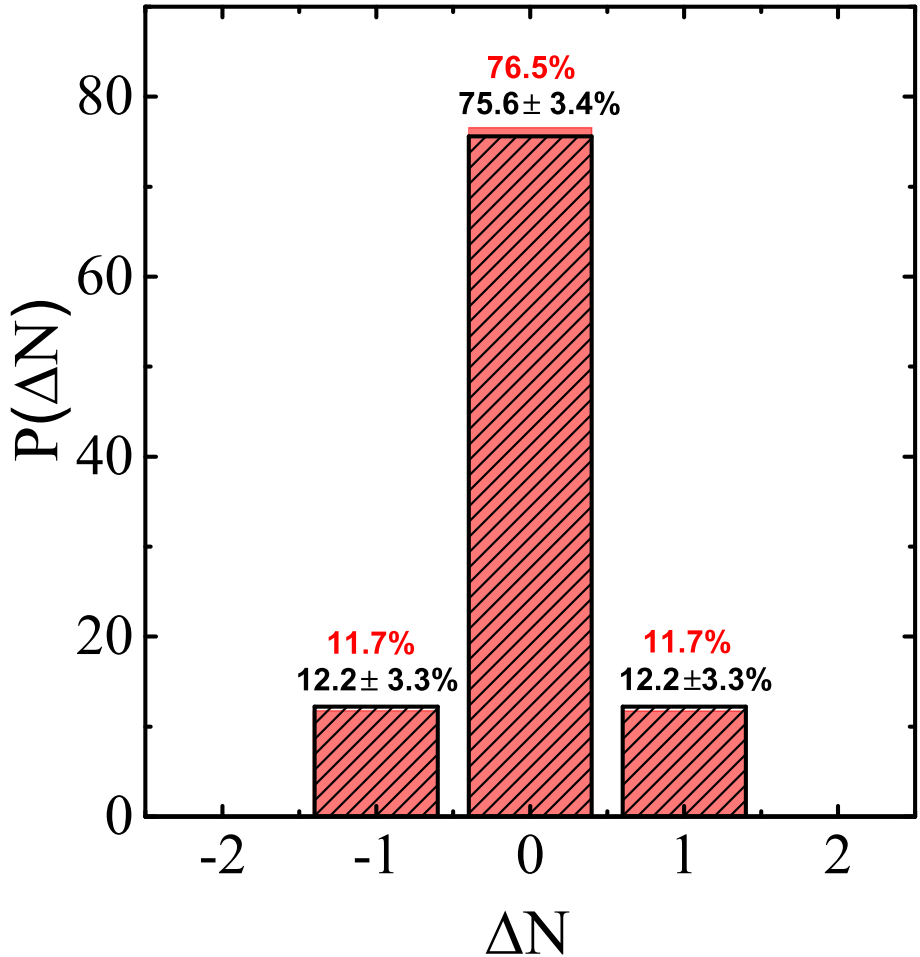}
\caption{Distribution  $P(\Delta N)$ of the spectral shift $\Delta N =N(\nu) - \tilde N(\nu)$  (black bars) measured for the microwave networks with preserved time reversal symmetry under the edge switch. The experimental results are compared with the numerical ones \cite{Aizenman2017} (red bars). } \label{Fig3}
\end{figure}

\subsection{Networks simulating quantum graphs with partially violated time-reversal invariance}

In order to investigate the edge transformations in the systems with partially violated time reversal invariance we used microwave networks with a microwave circulator Aerotek D93-1FFF (see Fig.~4).  The time violation was induced with a microwave circulator with low insertion loss which operates in the frequency range from $0.8-2.5$~GHz. A microwave circulator is a non-reciprocal three-port passive device. A wave entering the circulator through port 1, 2 or 3 exits at port 2, 3, or 1, respectively, as illustrated schematically in  Fig.~4.  The networks were connected to the Agilent E8364B vector network analyzer (VNA) via the HP 85133-616 microwave flexible cable. The resonances $\nu_n$ of the microwave networks were evaluated by measuring a single-port  scattering matrix $S(\nu)$  as  a function of microwave frequency $\nu$.

\begin{figure}
\includegraphics[width=18.cm, angle=0]{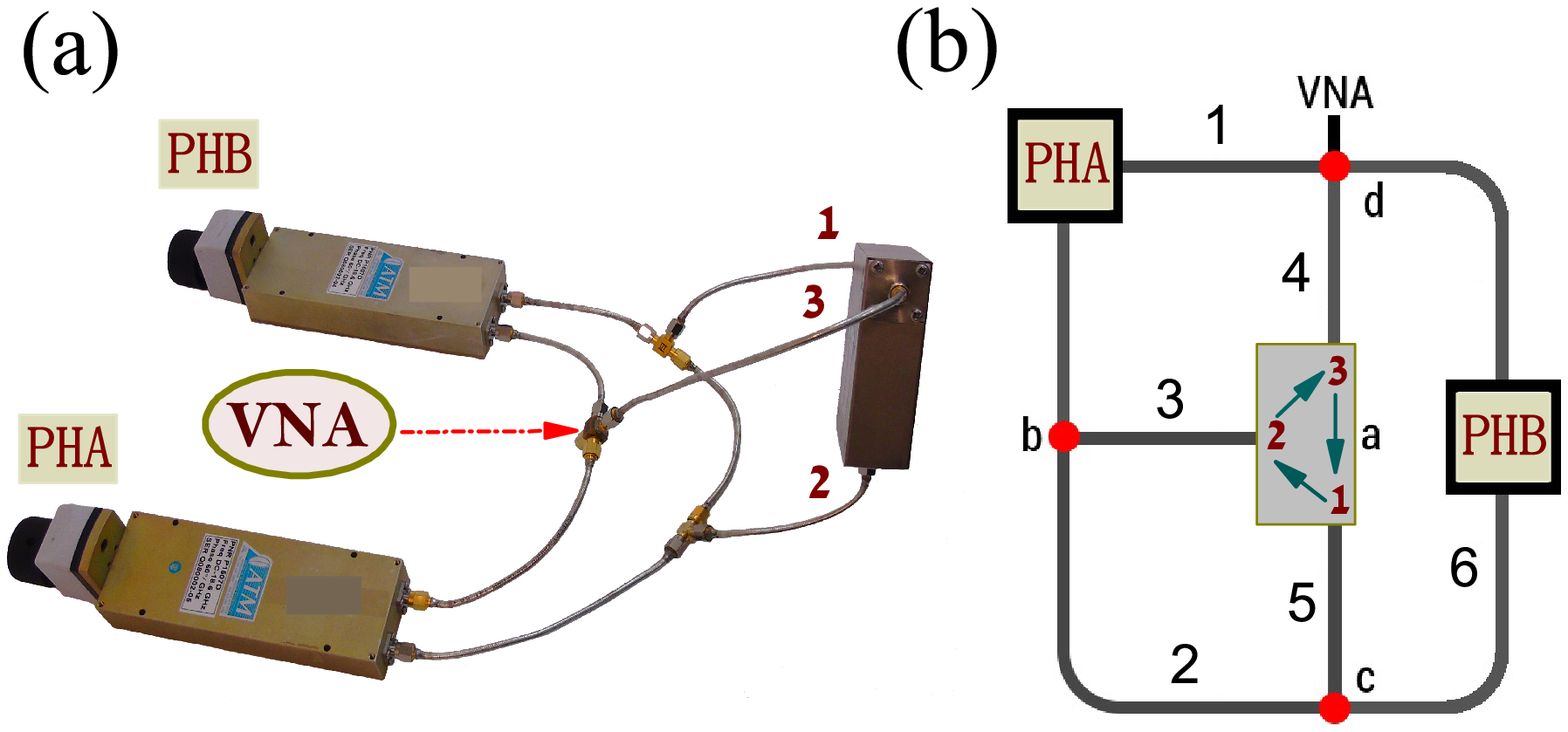}
\caption{ (a) Photo of a microwave network  simulating a quantum graph with violated time reversal symmetry. The network contains two phase shifters PHA and PHB and a three port microwave circulator.  (b) The scheme of the microwave network. The green arrows in the circulator show the directions of the energy flow. The vector network analyzer (VNA) was connected to the vertex $d$ of the network.} \label{Fig4}
\end{figure}

\begin{figure}
\includegraphics[width=12.cm, angle=0]{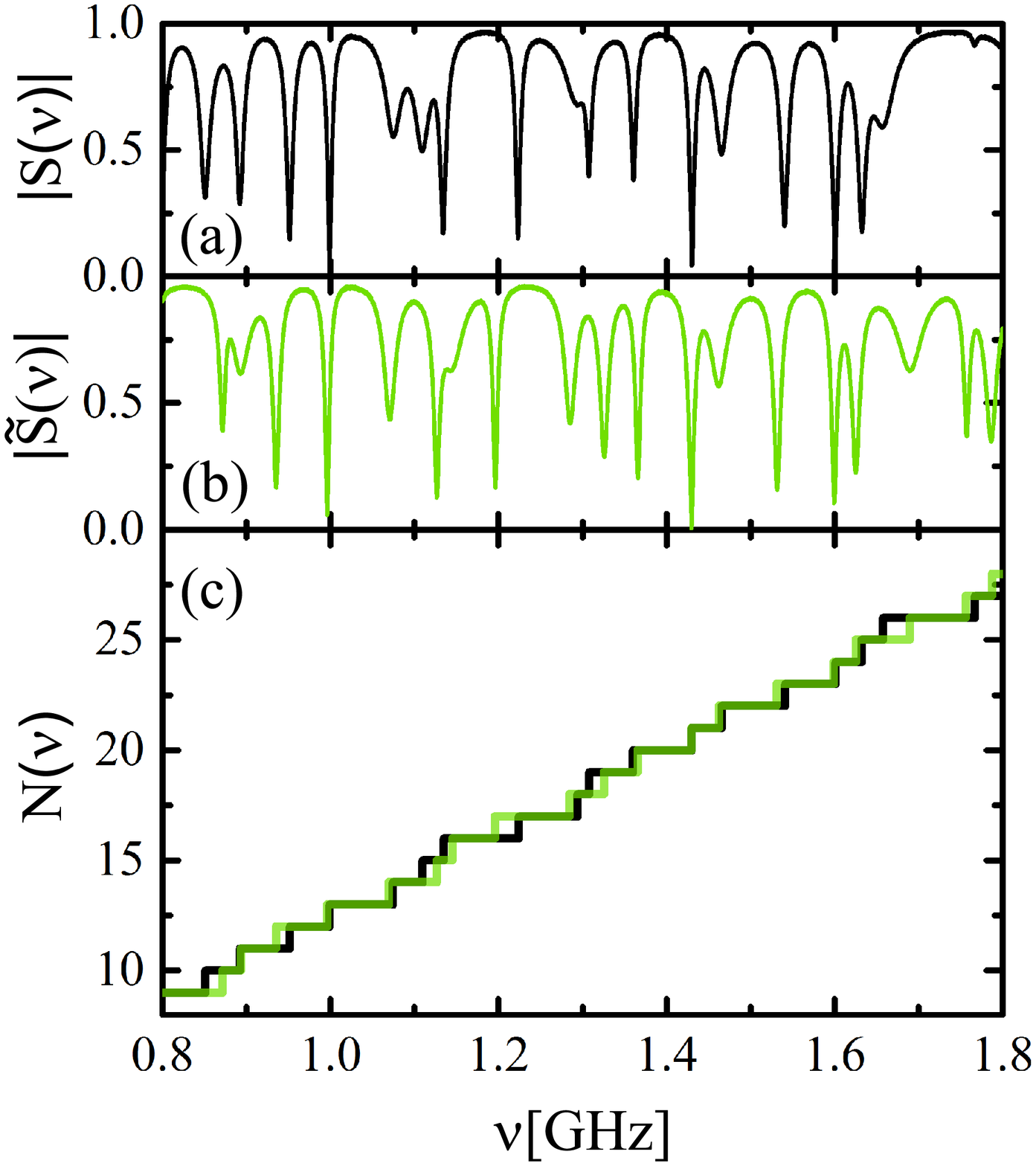}
\caption{(a) An example of the modulus of the scattering matrix $|S(\nu)|$ measured for the network with violated time reversal symmetry before the edge switch transformation in the frequency range 0.8-1.8 GHz. (b) The modulus of the scattering matrix  $|\tilde S(\nu)|$  measured after the edge switch transformation. (c)  Comparison of the counting functions $N(\nu)$ (black lines) and $\tilde N(\nu)$ (green lines) of the ``original'' and ``switched'' configurations for the microwave network with violated time reversal symmetry.} \label{Fig5}
\end{figure}

Similarly to the previous case the total optical length of the network  $L_{opt}=2.918\pm 0.010$~m was kept constant by increasing the length of one bond and decreasing the length of another one by the same amount using the phase shifters (Advanced Technical Materials PNR P1507D).
Therefore, while the length of the edge $1$ was enlarged by increasing the phase of the phase shifter $PHA$  from $0\degree$ to $42\degree$ in seven $6\degree$ steps, the phase of the second phase shifter $PHB$ in the edge $6$ was simultaneously decreased from $42\degree$ to $0\degree$ in the same number of steps. Due to microwave circulator's characteristic the scattering matrix $S(\nu)$ of the microwave networks was measured for all phase shifters positions in the frequency window $\nu= 0.8-2.5$ GHz.  The switch operation was realized by interchanging the cables $3$ and $2$ but keeping their connectivity in the vertex $b$. The lengths of the edges $3$ and $2$ were accordingly $ 0.170 \pm 0.001$ m and  $0.327 \pm 0.001$ m.
In this frequency window we were able to identify first $33-34$ resonances.  The number of the measured resonances  was in agreement with the one predicted by the Weyl's formula $N_{av}(\nu)=2L_{opt}\nu/c$, which in the frequency range 0.8-2.5 GHz yields $\simeq 33$ resonances. In Fig.~5(a) and Fig.~5(b) we present in the frequency range 0.8-1.8 GHz the examples of the moduli of the scattering matrices $|S(\nu)|$ (black full line) and $|\tilde S(\nu)|$ (red full line) measured for the network before and after the edge switch transformation. In Fig.~5(c) we compare the counting functions $N(\nu)$ (black lines) and $\tilde N(\nu)$ (red lines)  of the ``original'' and ``switched'' configurations.

Fig.~6 shows that although our microwave networks  simulating quantum graphs with partially violated time-reversal invariance are relatively simple, nevertheless the level spacing distribution $P(s)$ (red line) calculated for an ensemble of 33 networks (1108 resonances)  is close to the  GUE prediction (black full line) in the random matrix theory. In Fig.~6 we compare our experimental results with the analytical distribution  $P(s,\xi)$ (black broken line) describing the transition from GOE ($\xi=0$) to GUE ($\xi \rightarrow \infty$) \cite{Lenz1992,Dietz2019}
\begin{equation}
 P(s,\xi) =  \sqrt{\frac{2+\xi^2}{2}}sc^2(\xi)\erf \left (\frac{sc(\xi)}{\xi}\right )e^{\frac{-s^2c^2(\xi)}{2}}   \,,\label{eq:PGUE}
\end{equation}
where $ c(\xi)=\sqrt{\pi[(2+\xi^2)/4]}\{1-(2/\pi)[\tan^{-1}(\xi/\sqrt{2})-[\sqrt{2}\xi/(2+\xi^2)]\}$ and $\erf(x)$ is the error function. We determined $\xi=1.0 \pm 0.2$ by fitting the expression (\ref{eq:PGUE}) to the experimental distribution $P(s)$. Fig.6 shows that already for  $\xi=1.0 \pm 0.2$ the analytical distribution  $P(s,\xi)$ is close to the GUE one.

In the inset in Fig.~6 we show the level spacing distribution $P(s)$ (light green line) calculated  numerically for the GUE graphs using the formalism of the scattering matrix developed in \cite{Kottos1999}. In the numerical calculations the microwave circulator was replaced by the vertex $a$.  Then, the magnetic vector potential $A$  which breaks the time reversal symmetry \cite{Kottos1999} was applied on all bonds of the network presented in Fig.~4(b). The numerical calculations were performed for 40 different configurations of the graphs before and after the edge switch transformation. In order to obtain better level statistics our calculations were extended to the frequency range $0-10$ GHz. All together we identified  $5960$ resonances. The comparison of the numerical results with the GOE (grey line) and GUE (black line) predictions clearly shows that we deal with the system with violated time-reversal invariance.

\begin{figure}
\includegraphics[width=12.cm, angle=0]{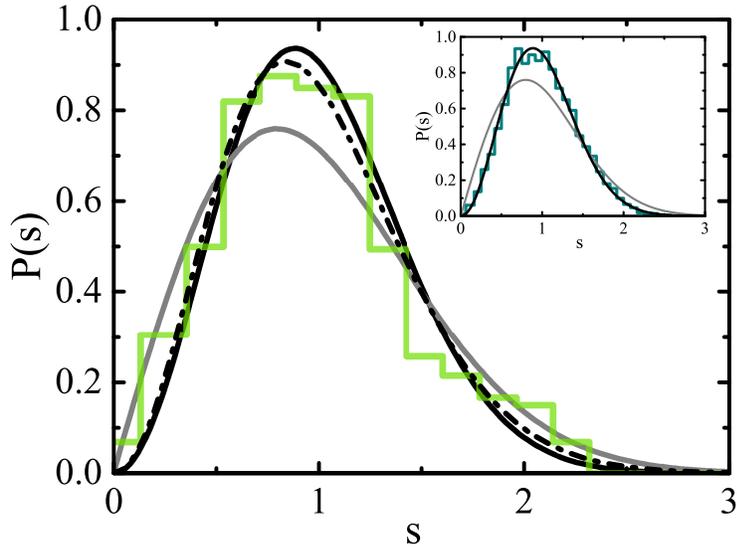}
\caption{The experimental level spacing distribution $P(s)$ (green line) obtained for an ensemble of $33$ microwave networks simulating quantum graphs with partially violated time-reversal invariance. The GOE and GUE predictions in the random matrix theory are shown by  full grey and black lines, respectively. The analytical distribution  $P(s,\xi)$ (black broken line) describing the transition from GOE to GUE was calculated for $\xi=1.0$.
The inset shows the  level spacing distribution $P(s)$ (dark green line) calculated for an ensemble of $40$ quantum graphs with violated time-reversal invariance. The GOE and GUE predictions in the random matrix theory are shown by  full grey and black lines, respectively.} \label{Fig6}
\end{figure}

\begin{figure}
\includegraphics[width=15.cm, angle=0]{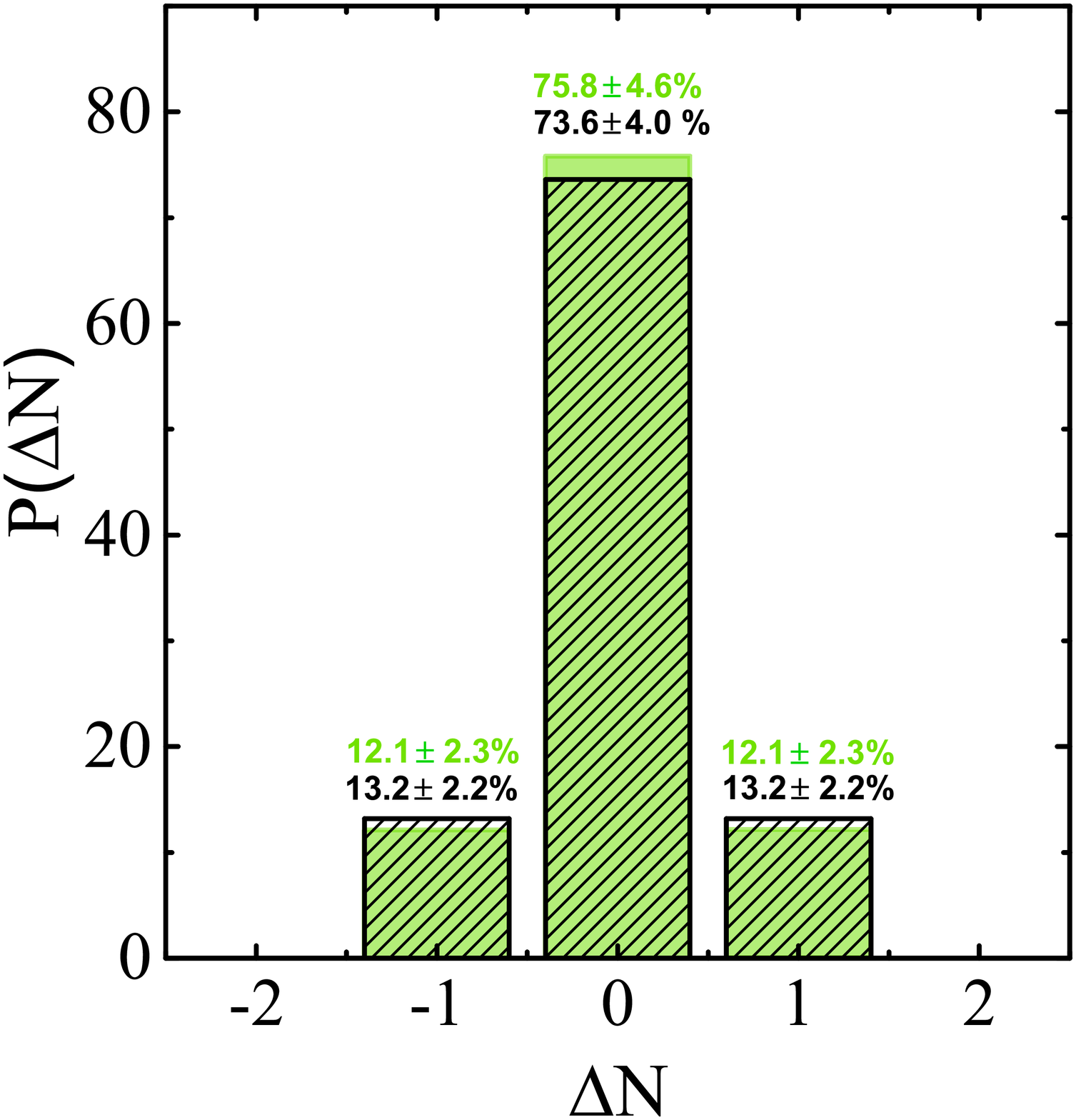}
\caption{Distribution  $P(\Delta N)$ of the spectral shift $\Delta N =N(\nu) - \tilde N(\nu)$ measured for the microwave networks with partially violated time reversal symmetry under the edge switch (black bars). The numerical results for the graphs with violated time-reversal invariance are  marked by the green bars. } \label{Fig7}
\end{figure}

In Fig.~7 we show the experimental distribution $P(\Delta N)$  of the spectral shift $\Delta N =N(\nu) - \tilde N(\nu)$ (black bars) obtained for the switch operation in the network simulating a quantum graph with partially violated time-reversal invariance. The results were obtained using the spectra of 16 network configurations. The uncertainty of the spectral shift $P(\Delta N)$ was calculated as a standard deviation error obtained from 8 pairs of the measured spectra. The experimental distribution of the spectral shift $P(\Delta N)$  is compared to the numerical one (green bars) calculated for the GUE graphs. The agreement between the experimental results and the numerical ones is good. Fig.~7 clearly shows that we deal with the spectra which are level-1 interlaced. It is worth to point out that the distributions $P(\Delta N)$ obtained for the systems with violated or partially violated time-reversal symmetry are very similar to the one obtained for the networks with preserved time reversal symmetry (see Fig.~3).

\section{Conclusions}

Using microwave networks with preserved and partially violated time-reversal invariance we showed experimentally that under the edge switch operation the spectra are level-1 interlaced with that of the original one, i.e., $\nu_{n-r}\leq \tilde \nu_{n}\leq \nu_{n+r}$, where $r=1$. For both symmetry classes, GOE and GUE,  the distributions $P(\Delta N)$  of the spectral shift $\Delta N =N(\nu) - \tilde N(\nu)$ are close to each other.
 Therefore, our results suggest that the interlaced properties of spectra of graphs and  microwave networks induced by the edge switch transformation may not be used to differentiate between the systems with GOE and GUE symmetries.
 However, the departure from the level-1 interlaced properties of the analyzed spectra can be used as an indicator of missing resonances.

Acknowledgments. This work was supported in part by the National Science Centre, Poland, Grants Nos. UMO-2016/23/B/ST2/03979 and UMO-2018/30/Q/ST2/00324.

\end{document}